\documentclass{article}

\usepackage[english]{babel}
\usepackage{latexsym}
\usepackage[usenames]{color}

\begin{document}

\title{Decision Taking versus Action Determination%
}
\author{%
Jan A. Bergstra\\
{\small  Section Theory of Computer Science,
Informatics Institute,} \\
{\small Faculty of Science,
University of Amsterdam, The Netherlands.}%
\thanks{Author's email address: {\tt j.a.bergstra@uva.nl}. This paper was written in connection with the ceremony at the first of June 2012 in the ``Aula'' of the University of Amsterdam highlighting the retirement of Frans Groen as a professor of informatics with the Informatics Institute of the Faculty of Sciences of the  University of Amsterdam.}
\date{}
}

\maketitle

\begin{abstract}
\noindent Decision taking is discussed in the context of the role it may play for various types of agents, and it is contrasted with action determination. Some remarks are made about the role of decision taking and action determination in the ongoing debate concerning the reverse polder development of the hertogin Hedwige polder.
\end{abstract}

\section{Introduction}\label{sec:Intro}
In \cite{Bergstra2011a} I have proposed an explanation of decision taking (DT) and the way it is embedded in and differs from decision making (DM). In \cite{Bergstra2012a} the distinction between decision taking and various forms of voting and promising is elaborated in the context of an investigation of the concept of decision taking as a service.%
\footnote{In \cite{Bergstra2011a} it is found that at least in principle decision taking can be outtasked (see \cite{BDV2011c}), but that it cannot be outsourced, thus contradicting a suggestion to that end made in \cite{Valdman2010}.}  
In this paper I will make a modest attempt to link the work of \cite{Bergstra2011a} and \cite{Bergstra2012a} with some topics of Frans Groen's research. Frans Groen's work mostly had a focus on methods and techniques that an artificial agent can use in order to choose an appropriate behavior in a physical environment. For instance in \cite{JansenMHG2005} colour vision is used in such a way that it may improve the obstacle avoidance capability of an artificially controlled car, an obvious necessity for automated off-road driving, and in \cite{DorpGroen2003} radar is used to improve upon the detection of walking humans above the capabilities that human agents have to that end. The technical focus of his research work has been on a wide range of sensor systems and on corresponding processing methods for sensor data. 

Each of these techniques can be, and ultimately will be, incorporated in autonomous systems governed by software agents. This perspective was a fundamental theme for Frans Groen's work for the last  20 years. Software agents may operate individually or in groups. Software agents need to trigger actions and to make choices and may be involved in collective decision making such as described in \cite{GroenSKP2007}. These choice processes are comparable to natural human action taking place in real time, see e.g.~\cite{MahalelZK1985}.

\subsection{Decisions are scarce}
Now I hold that when operating in real time both artificial and human agents don't engage in decision taking in spite of the fact that they may use sensor data and background information to compute what action to perform next. By consequence I hold that each of the contributions of Frans Groen to the theory, the technology, and the practice of autonomous systems can be profitably incorporated in applications without playing a role in decision taking. Decision taking may even be considered a redundant concept for
artificial autonomous systems. Further decision taking is very difficult for groups of human agents. In addition I hold that animals are incapable of decision taking. Together this means that as a fraction of choice processes on future behavior decision taking is relatively scarce.

These positions can be defended only on the basis of a sufficiently exotic definition of what it means to take a decision. In \cite{Bergstra2011a} I have argued that (i) a decision is an act of decision taking, performed by an agent (decision taker), operating in a specified role, having explicit intentions, and equipped with an explicit expectation of how its decision will contribute to a realization of these intentions, (ii) a decision produces a decision outcome, which is a tangible piece of information, (iii) the decision outcome may trigger agents in its scope to put it into effect, thus leading to the consequences of the decision outcome, (iv) decision taking constitutes a final phase of decision making, (v) decision taking  plausibly involves carrying out a protocol, the preparation of which is a task of the decision making process, (vi) determination of the content of the decision outcome and of parts of it belongs to decision making (and in particular to what is termed decision preparation in  \cite{Bergstra2011a}) rather than to decision taking.%
\footnote{An example of a protocol element of a decision taking thread occurs  when selling a home (that is just before transferring economic ownership in a formal session): (i) one needs to check that the property has been properly insured by the buyer, (ii) one needs to check having available all keys, (iii) one needs to have available all current metering data concerning water and various forms of energy, (iv) the property has been brought in the required state, (v) one needs to have passports or other means of identification available for all sellers, or otherwise have transferred their representation to someone else who will attend the session, (vi) a bank account has been provided to the solicitor which can accommodate the sum that will be transferred once the legal ownership has been transferred as well.}

The garbage can model of decision making of \cite{CohenMO1972} fits well in this view of decision taking. In that model decision making produces plans that may constitute candidate solutions to forthcoming problems, while decision taking singles out and activates candidate solutions that are considered proper solutions for actual problems, at appropriate moments.

\subsection{Action determination, a universal behavioral pattern}
If an agent  engages in an action on the basis of some reflection about past and recent data as 
well as a portfolio of rules of conduct  it can be said that the agent has determined a plan or an action for immediate effectuation. Determining what to do next is a frequent process for any agent, including artificial agents. I will speak of action determination. For a human agent it is common to say that the agent makes up his or her mind concerning an action to come. 

Action determination can be phrased in different ways: choosing an option, where the option is chosen from a menu of actions; or selecting an alternative from a set of alternatives. Action determination leads to the determination of an action. The action that is determined  may also be referred to as the determination outcome.%
\footnote{Unlike a decision outcome an action determination outcome is not primarily a piece of information. Perhaps a more systematic terminology is obtained when ``action determining'' is used instead of ``action determination''. Then it may be said that a determination is an act of action determining, just as a decision is an act of deciding (decision taking).} 
Unlike with decision taking as conceived in \cite{Bergstra2011a}, the consequences of natural decision taking are in general not brought about through the intermediate stage of a  decision outcome.

For a choice  process  proceeding in real time (such as the driving activity of \cite{MahalelZK1985}) the phrase ``natural decision making'' has been coined (see e.g. \cite{ThwaitesWilliams2006} for an application or \cite{ZsambokKlein1997} for an introduction of the notion). The phrase ``natural decision taking'' provides better consistency with the terminology of \cite{Bergstra2011a}. But I prefer to speak of action determination rather than of natural decision taking, because action determination suitably highlights the real time character of the choice process involved.%
\footnote{Process algebra (see \cite{BaetenBastenReniers2009}) can be understood as a theory of action determination with a primary focus on (i) immediate effects of actions in terms of process interaction (communication), (ii) external visibility of actions (abstraction), (iii) enabling of actions (encapsulation), and (iv) preferences between actions (priorities).}

Some examples of the use of ``action determination'' may be helpful: (i) a car driving agent who notices the approach of  a traffic light that has just turned red must determine whether to stop (and if so whether to stop abruptly (and if so whether stop before the white demarcation line linked with the traffic light or to tolerate passing over that line some fraction of a car length), or to stop gently), or to proceed (and if so, at an accelerated speed, at a constant speed or at a lower speed). (ii) When no
service is being offered in a shop to an agent acting in the role of a candidate customer the agent must determine whether to wait 
(and if so, to perform that determination repeatedly until either the agent has been served or in determines that the waiting must come to an end), or to leave the shop.

\subsection{Real time action determination}
Action determination mainly takes place in real time. Real time requirements often  preclude the construction of an information carrying outcome as an intermediate stage for action determination.

Sensor data from devices monitoring the physical environment can play a role  in decision preparation, probably after sampling and subsequent statistical processing. Instead during 
decision taking, which is a relatively slow process heading for an outcome consisting of  a piece of data, the usage of real time collected sensor data is less plausible. The process is too slow for making relevant use of real time measurements and too fast for systematic sampling with subsequent statistical evaluation. Nevertheless, real time data from social media may be taken into account in a decision taking protocol, for instance by blocking a decision if there is a large number of opponents online. 

Action determination in real time can be performed rationally if it is based on some theory about what is to be optimized. Limitations of an agent's computational capacity will give rise to bounded rationality in real time action determination.

\subsection{Action determination and social choice}
Action determination need not involve decision taking but in some cases decision taking may be used as a way to 
achieve  action determination. Action determination can be performed on the
basis of rational rules in which case one may speak of  rational action determination. 

Action determination may also be implemented by means of social choice. Action determination is common to all forms of agents and also it is also a common process for groups of humans.
For all types of agents action determination seems to take place more frequently than decision taking. Phrased differently, action determination is less scarce than decision taking.

\section{Mechanical aspects of decision taking}
Following \cite{Bergstra2011a} it is essential for a decision that it produces a decision outcome as an intermediate product, 
which by itself, and solely based on the role of the decision taker, not on the decision taker's subsequent actions, 
influences its environment to enact consequences that are in conformance with the  decision taker's 
expectations and from which the decision taker expects a positive contribution to the realization of its objectives. 

As was mentioned in \cite{Bergstra2011a} I consider it implausible that an animal agent produces an intermediate piece of information encoding its intentions and expectations, and I doubt that large groups of humans can
entertain intentions or have expectations, from which I infer that decision taking is less plausible for large human groups. the case of small human groups is different. I assume that such groups can have intentions and objectives and that it makes perfect sense to think of their decision taking activity.

\subsection{Software agents choose without deciding}
For software agents the argument against decision taking being a prominent behavioral pattern is slightly different than the arguments given for animals and for large human groups. Taking decisions requires an explicit maintenance 
of intentions and expectations, and a setting where its own actions as well as that of other agents are based upon inspection of a portfolio of decision outcomes. Explicit maintenance of decision outcomes as information objects is certainly an option  for the design of a software agent, but it is definitely not an essential feature for artificially intelligent behavior. 

Further a decision taker must operate in the context of a role which provides weight to its 
decision outcomes for other agents. That
an artificial agent acts within a role with authority is hard to imagine, unless it impersonates a human agent acting within such a role.%
\footnote{I am limiting attention to artificial agents  that are perceived as being artificial by all other agents which   they are interacting with.}

For these reasons I consider decision taking to be an advanced feature for artificial agents, though not an inaccessible feature. Action determination or making choices from menus of alternatives, which is often viewed as the core of decision making (a viewpoint that I do not share), however constitutes an essential feature of artificial agent behavior.

\subsection{Software agent control: a multi-scale phenomenon}
\label{Msp}
The behavioral control of a software agent is a phenomenon which requires a multi-scale approach: (i) at the hardware level some controls of  a software controlled agent constitute the lowest scale which is out of reach of the software, (ii) at the software level action determination may be programmed by means of simple algorithms, and (iii) at a higher level (scale) action determination may involve explicit reasoning, (iv) action determination may depend on the generation of intermediate pieces of information thus featuring artificial decision taking, (v) at the largest (highest) scale of its functionality a software agent interact with other software agents in order to participate in artificial social choice mechanisms, and (vi) at yet another scale a human engineer takes decisions about the deployment of software for an embedded software agent (and so on).

I have come to suspect that several concepts in computing, among which the notion of a computer program, are 
essentially multi-scale notions in the sense that an account of those concepts in terms of a single scale turns out to be defective invariably. An intriguing example of a multi-scale concept is money. To see this one may notice that at the scale of a single household it is clear that there may be a lack of money, but at the level of an entire economy speaking of that kind of shortage makes much less sense.

\subsection{Multi-threading and short-circuit logic}
I consider it to be a realistic assumption that a decision taker is involved in a plurality of decision taking threads simultaneously. That form of concurrency can be described adequately with strategic interleaving as specified in \cite{BergstraMiddelburg2007}. If a protocol for  taking a particular decision is applied that application may amount to the effectuation of an instruction sequence (e.g.~as in \cite{BergstraLoots2002a,BergstraMiddelburg2012}) in a suitable execution architecture (see \cite{BergstraPonse2007a}), involving  the evaluation of composite conditions (as in
\cite{BergstraPonse2011a}), thus generating a sequential thread. 

The evaluation of composite conditions in real time may bring with it a dynamic setting where for evaluating a single composite condition the same boolean value needs to be successively evaluated several times and takes different values. That situation leads to so called reactive valuations as 
proposed in the short-circuit logic of  \cite{BergstraPonse2012a}. 

An important consequence of a decision taking agent  being engaged in a plurality of decision taking threads concurrently is that a serious failure in one thread may induce a premature exit from other threads as a side-effect. This typically happen with agents in a political role and who may loose support concerning their handling of a particular case thereby negatively impacting on their competences in other areas of decision taking.

\subsection{Protocols and modularization}
I view decision taking  as a form of modularization of organizational behavior. Stated differently: decision taking is an organizational pattern or feature. As a feature it may have different degrees of visibility in an organization's operation. For instance, the preparation for the so-called ``instellingsaccreditatie''
(institutional accreditation), at the time of writing ongoing for the University of Amsterdam, requires an effort to increase the amount of decision taking that University's processes at many levels. 

Changing  an organization by installing new patterns of decision making and decision taking can only be done if a prepared portfolio of decision taking protocols is available. 
The design of decision taking protocols in terms of instruction sequences making use of short-cicuit logic, and of  strategic interleavings for threads resulting from the effectuation of such protocols, is a challenge for forthcoming research. This task is especially interesting in connection with human decision taking. For instance when buying a second hand car it may be practical to have a prepared protocol at hand for checking all relevant matters before signing.%
\footnote{Signing a contract for buying a car certainly qualifies as taking a decision, with the contract playing the role of the decision outcome.}

While decision taking can be understood as an optional architectural feature for organizational design, action determination is unavoidable and to propose its understanding as an aspect of a software agent's software architecture would be misleading.

\section{Group decisions, a controversial example}
\label{HP}
Decision taking and social choice plays a central role in the behavior of human groups. For instance regarding certain themes in Dutch politics it is essential to be able to analyze the situation in terms of social choice,  decision making, and decision taking. An interesting case is the long standing debate about whether or not to return the hertogin Hedwige polder to the influence of tidal movements in the Westerschelde. I will look into that case in some detail.

\subsection{Reverse polder development of the Hedwige Polder}
For the plan of this paper it would be preferable to use an example which (i) contrasts natural decision taking as mentioned above that  makes use of real time sensor data with decision taking as meant in \cite{Bergstra2011a}, and which (ii) at the same time illustrates the  contrast between decision taking and events of social choice. Unfortunately the example discussed is rather skewed towards the second aspect, but its timeliness makes up for that disadvantage.

The case that I will discuss in some detail concerns a long standing policy issue between Belgium and the Netherlands: the accessibility of the port of Antwerp. Nowadays the matter is significantly complicated by three aspects: (i)  the need to accommodate larger ships, especially container carriers, (ii) the need to take ecological consequences into account much more seriously then before, and (iii) the abundant use of social computing that produces immediate impact figures on any decision taker's popularity.

In \cite{Floor2009} and \cite{Tilburg2010}  meticulous accounts are presented regarding the remarkable number of actions, choices, decisions, and events of social choice and action determination that have  taken place since the so-called second improvement ({\em tweede uitdieping}) of the fairway to Antwerp has been requested by the Belgian government some 20 years ago. The key issue is that improving the accessibility of the Westerschelde for larger ships leads to ecological costs which can, and according to important voices, must be compensated for by reverse polder development ({\em ontpolderen} in Dutch), that is moving an existing polder outside the dykes which are protecting it against the open waters  of the Westerschelde, in order to obtain a specific type of wetlands considered to be of essential ecological value, thus preserving the volume of such wetlands in the Schelde estuary. This is technically achieved by placing new dykes behind the polder and subsequently removing the dykes in front of it at least partially. The hertogin Hedwige polder in Zeeuws Vlaanderen has become a focus of attention for this matter. Local farmers and land owners strongly oppose the process of inundating a polder currently providing valuable farmland, mainly on emotional grounds, and claiming that returning a polder to the sea goes against a tradition of centuries as well as against local and even regional economic interests.

A cascade of so-called decisions has produced a state in which it has become almost impossible to understand the merits of reverse polder development relative to the merits of keeping the polders and their usage unchanged. At the time of writing this paper it is unknown what the fate of the Hedwige polder will be. It is probably difficult to predict what will happen for all agents involved. The matter is obviously problematic for the Dutch government and it may lead to a serious conflict with the port authority in Antwerp. Such conflicts have in fact existed for centuries in the past, only to be resolved by Napoleon in 1792. A treaty dating back as a long as 1839 about enduring guarantees for the accessibility of the harbor of Antwerp and its maintenance still impacts on these matters. 

\subsection{Justification of some candidate decision outcomes}
Regarding this case I have come to several conclusions in connection with decision taking and action determination, 
in particular  regarding the possible justification of some candidate decision outcomes. These are my personal and subjective viewpoints in a case where other persons have chosen different positions about various aspects of the matter.
\begin{enumerate}
\item For a clear analysis of the political situation concerning the Hedwige polder and its past it is useful to distinguish action, choice, and decision, as well as to maintain a distinction between decision taking, decision making, and  social choice, and to distinguish decision outcome from social choice outcome. 

\item A recent development is that some forms of social choice can be realized in real time by means of statistical and automatic analysis of social media traffic.  This is ``socio-technical sensor technology''$\!$, a part of  ``socio-technical informatics'' (or ``socio-technical informaticology''). Socio-tech-nical informatics might be productively 
grouped together as a theme with classical sensor technology that has been the focus of work of Frans Groen. Results of socio-technical sensing, which in The Netherlands
often take the form of quite arbitrary and disputable polls about hypothetical voting outcomes for the Dutch parliament, seem to impact decision processes and events of social choice aimed at real time action determination.%
\footnote{In \cite{DorpGroen2003} radar is used to see walking persons through a wall, an instance of observation outside the range of capabilities of human observers. That issue may not be so distant after all from measuring a persons preferences about a policy issue by asking questions about seemingly unrelated matters.}
\item It seems to be a hidden assumption made by many actors relevant to this issue  that full or partial reverse polder development of the Hedwige polder can only take place after a decision to do so has been taken, and has not been itself reversed in subsequent legal action initiated by its opponents. This assumption is wrong if the concept of decision of \cite{Bergstra2011a} is used. It is wrong because it might be the case that only events of social choice come into play which may qualify as action determinations but do not qualify as decisions.

The occurrence of an event of social choice does not imply the existence of an agent with corresponding intentions and expectations, and for that reason an event of social choice may not count as a decision. Human groups tend to produce instances and corresponding outcomes of social choice rather than decisions and corresponding 
decision outcomes. For human groups (like the Dutch parliament) taking decisions is difficult.
\item It can be maintained, however,  that action determination  will be required as a precondition for each form of reverse polder development in the Netherlands. The relevant action determination phase may, however, comprise no more than one or more events of social choice each failing to qualify as decision taking.
\item Taking a decision that amounts to reverse polder development of the entire Hedwige polder is justified at this stage.%
\footnote{Some recent results of social choice that have been produced by the Dutch parliament in this matter have an interesting complexity from a logical point of view. For instance the following package: (i) to start with improving the waterway, (ii) and at the same time to prepare for reverse development of the Hedwige polder, and (iii) concurrently to look for alternatives  which allow to preserve the Hedwige polder, which (iv) are given priority over reverse polder development when considered sufficiently attractive. Abstraction from the many internal steps involved in the threads of these activities is an implicit assumption when working at the level of abstraction of a legislator. A process theory sufficiently powerful to ``specify'' the plan prescribed by this outcome of social choice must  combine concurrency, composition of alternatives, sequential composition, conditions, priorities, and abstraction. Such process theories are rare, an example of a process model which meets these requirements, so-called orthogonal bisimulation semantics,  has been given in \cite{BergstraPonseZ2003a}.}
If the government had been able to insist on taking decisions (that is ministers taking decisions strictly corresponding to their own intentions and expectations  and getting the resulting decision outcomes subsequently ratified in parliament), rather than to have their intentions cluttered up by modifications proposed by MP's with various backgrounds, then a decision outcome of this kind would probably have been obtained at this stage already.

\item Thus I consider a decision outcome that amounts to reverse polder development of the entire Hedwige polder justified. Let full rpdHP abbreviate that (potential) decision outcome, and let partial rpdHP denote an outcome by which  a part of the Hedwige polder is preserved. Let no-rpdHP abbreviate the decision outcome that no full or partial reverse polder development of the Hedwige polder is carried out and no further  measures are taken to compensate for that. Now I wish to state that:
\begin{enumerate}
\item I already had a prior preference for full rdpHP over no-rdpHP and over partial rdpHP based on a ``green sentiment'' before reading a selection of the available documentation about the matter.
\item I consider the presence (in my mind) of a prior preference for one potential decision outcome over another potential decision outcome not to be in contradiction with the imperative of impartial analysis. The case seems to be comparable to the existence of prior odds in subjective probability theory and Bayesian statistics. By becoming increasingly aware of pieces of information one's subjective justification of  a particular decision outcome as compared to a different candidate decision outcome may be adapted in successive stages. The presence of a prior preference for a possible decision outcome is to be expected if one accepts a method of subjective justification of potential decision outcomes. 
\item  I do not  understand in detail the entire chain of arguments, both legal and ecological, that leads to the claimed necessity of full or partial rdpHP; these arguments are quite technical and have a multi-disciplinary nature, in particular the expected negative ecological consequences of the ``tweede uitdieping'' are not so easy to grasp.%
\footnote{Ecosystem preservation seems to be a multi-scale concept (see \ref{Msp}), a feature that creates significant confusion.} The latter difficulty perhaps  indicates a shortcoming in the communication of the matter by the proponents of full or partial rdpHP.
\item There may be a speculative link of this theme  with sensor technology as follows: by performing a detailed investigation, probably technically supported with extensive use of  sensing devices,  of water and sand movements in the estuary of the Schelde, a method might conceivably be discovered which allows to make the fairway to Antwerp accessible for larger ships and at the same time to prevent degradation of the ecosystem. Mobile elements in open water, reacting on the presence of ships,  as well as dynamic and natural reconfiguration of tidal flows may be instrumental for such solutions. At present, however, such speculative options  cannot be brought forward as an argument against full or partial rpdHP.
\item  As an owner of a modest amount of farmland some 7 kilometers to the west of the Hedwige polder I may have a personal interest in the matter. However, I am not sure about  the implications of that connection; my stated preference deviates from the ZLTO%
\footnote{Zeeuwse Land en Tuinbouw Organisatie.} position to which I might supposedly be attracted in the mentioned capacity, and the implications of full or partial rpdHP on the value of farmland in the area are difficult to estimate, and so are the implications of choosing a position that deviates from that of the local agricultural sector which, like ZLTO,  opposes to
both full and partial rdpHP.

\end{enumerate}
\item Given the incredible costs of decision making in this matter, secretary Bleker's recent proposal for a partial reverse polder development of the Hedwige polder, proposed in combination with a range of other probably less problematic developments,  constitutes a sounds step of (true) decision preparation. 

Regarding that proposal it is amazing that an additional cost of some 150.000.000 Euro is considered acceptable in order to avoid the reverse polder development of some 2.000.000 square meters, thus protecting two thirds of the Hedwige polder against inundation. It is hardly conceivable that the opponents of reverse polder development would  consider spending that much money on the preservation of agricultural area if they could not force the state to pay this excessive sum. Rather than being dissatisfied with the result (if Bleker's plan came true), its opponents should be grateful for such a formidable investment made by the Dutch government to meet their complaints, if only partially. 

Bleker's proposal has subsequently been rejected in the Dutch parliament in May 2012  and the issue has been postponed until after the next elections thus leaving matters wide open again.

\end{enumerate}

\subsection{Decision taking as a service (DTaaS)}
In \cite{Bergstra2012a} I have outlined that in some cases decision taking can be offered as a service by a service provider to a customer who was in charge of the same kind of decisions before outtasking that part of its decision taking load. Perhaps it will be helpful to have the decision taking concerning the future of the  Hedwige polder outtasked from political functionaries to some external agents. 

The search for useful majorities in parliament has become infected with 
issues like the so-called Euro crisis which are very distant from finding convincing solutions to ecological matters that don't reach beyond Belgium and the Netherlands. By using DTaaS accidental  dependencies of DT concerning the Hedwige polder on unrelated issues may be prevented.%
\footnote{At the time of writing all EU related policy matters, including compliance with EU regulations on ecosystem preservation,   are overshadowed by the so-called Euro crisis. We move through ``Grexiting times''$\!$, with significant excitement generated by a possible forthcoming exit of Greece (called Grexit by some) from the Eurozone. The multi-scale aspect of money (see \ref{Msp}) seems to be one of the  causes of this excitement. Those who claim that Greece simply must get its budget in order (looking at Greece at the scale of an ordinary household) hardly understand those who claim that the Euro system at large is at stake (with Greece constituting no more than a temporary center and highlight  of problems that are much more widespread) and that artificial restrictions on the availability and usage of money need to be addressed with higher priority.}

\subsection{Decision preparation as a service (DPaaS)}
Using the equation DM = DT + DP (decision making = decision taking + decision preparation) from \cite{Bergstra2011a} it is also plausible to consider the service casting (see \cite{Bergstra2012a}) of decision preparation (DP): ``decision preparation as a service'' (DPaaS) as an option besides service casting of DT (DTaaS). 

The following can be remarked about DPaaS: (i) DP can be outtasked and in contrast with DT it can also be outsourced, 
(ii) DPaaS is very common, outtasking of DP involves the services of consultants of various kinds, outsourcing of DP  involves moving teams of specialists to other organizations from where they will provide their expertise wrapped in a service, (iii) if DPaaS is used in combination with DTaaS it seems very important to see to it that different, and independent, service providers are used, (iv) DPaaS in case it comes about through outsourcing of DP activity may lead to a provider lock in, (v) DPaaS has occurred in the Hedwige polder issue to the extent that participants of the DT process may  have lost contact with the argumentation used by the relevant DP professionals, (vi) the latter effect, if true, might explain a tendency which seems to exist to drift towards the use of legal arguments by various actors in the DT process.

\subsection{Some risks for prospective decision takers}
Two groups in Dutch politics have chosen the position of definite opponents of full or partial rpdHP: SP and PVV. Apparently their ``negative view''$\!$, which I consider to be remarkably shallow in its motivation and analysis of the issue, is currently very powerful in terms of its effects in the media and the polls. This influence  through various social media is so strong that for other political groups it is not without risk to field any actor who proposes tho opt in favor of full or partial rpdHP 
simply as a decision outcome compliant with his/her own points of view about the matter when stripped from its legal history.%
\footnote{The appearance of such actors is required if decisions and decision outcomes as meant in \cite{Bergstra2011a} are to be produced.}

The perception of this risk induces a preference for members of other political groups to view the whole issue as being implied by the legal positions obtained in the matter after various episodes of negotiation with the Belgian government and the EU. This preference also opens the door for a display of the anti EU mentality of SP and PVV in the context of this particular case. Making that connection is not very convincing because the negotiations about the fairway in the Westerschelde and its maintenance have been properly conducted.%
\footnote{As it seems SP and PVV make use of an interesting argument in order to preserve the consistency of their position: the introduction of a labeling of the individuals who  negotiated about these matters  in the last 15 years. These are supposed to have been urban intellectuals who displayed an insufficient grasp of the business logic of farming.}

Now it is likely that SP and PVV will not take responsibility for actually refusing full or partial rdpHP if it appears that their opponents can see to it that this policy of refusal (when effectuated) will be followed by an assignment of responsibility for its consequences. According to the opponents of SP and PVV these consequences amount to the effects of international legal action by Belgian authorities that will eventually force The Netherlands to act as promised and  to compensate in financial terms for significant and unnecessary delays in doing so.

Protected by the expectation that SP and PVV are unlikely to have government responsibility soon, they take the liberty to maintain disputable viewpoints regarding the future of the Hedwige polder. This mechanism must be challenged by their political opponents. 

\section{Concluding remarks}
Decision taking (as viewed in \cite{Bergstra2011a}) provides a perspective on the behavior of human agents operating alone or in small groups.  Decision taking has been contrasted with action determination. Both action determination and  decision taking need to make use of data, either serving as parameters for determination outcomes or decision outcomes,
the real time collection and processing of which has been the research focus of Frans Groen, to whom I want to express my gratitude  for many years of effective and pleasant cooperation in  ``het IvI''.

\end{document}